\documentstyle[psfig,twoside,fleqn,espcrc2]{article}

\newcommand{\AmS}{{\protect\the\textfont2
  A\kern-.1667em\lower.5ex\hbox{M}\kern-.125emS}}

\title{Topology of Boundary Surfaces in 3D Simplicial Gravity}

\author{H.S.Egawa\address{Department of Physics, Tokai University Hiratsuka, 
                          Kanagawa 259-12, Japan}
                 $^{,}$
                 \address{Theory Division, Institute of Particle and Nuclear 
                          Studies, KEK, High Energy Accelerator Research 
                          Organization Tsukuba, Ibaraki 305 , Japan}
        and 
        N.Tsuda$^{\,\, {\rm b}}$
        \thanks{supported by Research Fellowships of the Japan Society for 
                the Promotion of Science for Young Scientists.}
}
\begin{document}

\begin{abstract}
A model of simplicial quantum gravity in three dimensions(3D) was investigated 
numerically based on the technique of dynamical triangulation (DT).
We are concerned with the genus of surfaces appearing on boundaries 
(i.e., sections) of a 3D DT manifold with $S^{3}$ topology.
Evidence of a scaling behavior of the genus distributions of boundary 
surfaces has been found.
\end{abstract}

\maketitle

\section{Introduction}
Recent numerical results obtained by dynamical triangulation for 3D and 4D 
suggest the existence of scaling behavior: for example, the number of 
simplexes at geodesic distance $r$, the fractal dimension and the volume of 
boundaries\cite{3D_Scaling,4D_Scaling}.
Furthermore, analytical results of $4$D QG show scaling relations of the 
partition function\cite{4D_Scaling_A}.
The purpose here is to further explore the boundary surfaces of 3D DT mfd.
If the scaling of the boundary volume distribution in $3$D and $4$D makes sense, 
\footnote{In 2D the scaling behavior of the loop-length distributions (LLD) is 
well known\cite{LLD}.}
it is expected that the boundary, itself, has physical significance.
We are concerned with the genus of these boundaries 
(see Fig.\ref{fig:3D_Boundary}).

Furthermore, the model of 3D Euclidean QG with boundaries may be recognized 
as a model of quantum nucleation of the universe in $(2+1)$-dimensional 
quantum gravity.
The nucleation of the universe by the quantum tunneling may be described by 
going out of the Euclidean signature region to the Lorentzian signature 
region in a sense of the semiclassical approximation.
The nucleation of the universe can also be regarded as a topology-changing 
process in the sense that the universe undertakes a transition from the initial 
state with no boundary to the final state with nontrivial topology 
(see Fig.\ref{fig:3D_Boundary}: $0 \to d$ or $d \to d^{'}$).
We can treat the model in a full quantum way, which means that we 
sum up the fluctuations of the 3D metric ($g_{\mu\nu}$).
The process of quantum tunneling requires that all of the components of the 
extrinsic curvature vanish (i.e., {\it totally geodesic}).
Although in our numerical simulations we put no restriction on the boundary 
surfaces, it is possible to introduce an extrinsic curvature on the boundary 
surface as a physical restriction.

\section{Model}
We start with the Euclidean Einstein-Hilbert action,  
\begin{equation}
S_{EH} = \int d^3 x \sqrt{g} \left(\Lambda - \frac{1}{G}R \right), 
\end{equation}
where $\Lambda$ is the cosmological constant, and $G$ is Newton's constant 
of gravity.
We use the lattice action of the 3D model,
\begin{eqnarray} 
S(\kappa_{0}, \kappa_{3}) & = & - \kappa_{0} N_{0} + \kappa_{3} 
N_{3}\nonumber\\    
  & = & - \frac{2\pi}{G} N_{0} \nonumber \\
  &   & + \left( \Lambda' - \frac{1}{G} (2\pi - 6 \mbox{cos}^{-1}(\frac{1}{3}))
          \right) N_{3} \nonumber ,  
\end{eqnarray}
where $N_i$ denotes the total number of $i$-simplexes, $\Lambda' = c\Lambda$; 
$c$ is the unit volume and $\mbox{cos}^{-1}(\frac{1}{3})$ is the angle 
between two tetrahedra; $\kappa_{0}$ is proportional to the inverse of bare 
Newton's constant, and $\kappa_{3}$ corresponds to a lattice cosmological 
constant.
The partition function is  
\begin{equation}
Z(\kappa_{0}, \kappa_{3}) 
= \sum_{T(S^{3})} e^{-S(\kappa_{0}, \kappa_{3})}.
\end{equation}

\section{Numerical Analysis}
\begin{figure}
\centerline{\psfig{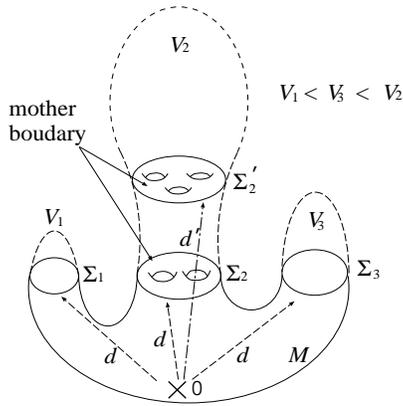}} 
\vspace{-1.0cm}
\caption
{
Schematic picture of boundary surfaces ($\Sigma_{1},\Sigma_{2}$ and 
$\Sigma_{3}$) at distance $d$ and a surface ($\Sigma_{2}'$) at distance 
$d'$ in 3D Euclidean space $M$ with $S^{3}$ topology.
The mother universe is defined as a boundary surface ($\Sigma_{2}$) with the 
largest tip volume ($V_{2}$), and the other surfaces are defined as the baby 
universes.
}
\label{fig:3D_Boundary}
\end{figure}
In order to discuss the scaling properties, we measured the genus 
distributions of the boundary surfaces with several geodesic distances. 
Suppose a 3D ball (3-ball) which is covered within $d$ steps from a 
reference 3-simplex in the 3D mfd with $S^{3}$ topology. 
Naively, the 3-ball has a boundary with spherical topology ($S^{2}$).
However, because of the branching of DT space, the boundary is not 
always simply-connected, and there usually appear many boundaries which 
consist of closed and orientable 2D surfaces with any 
topology and nontrivial structures, such as links or knots.
The boundary surfaces are divided into two classes:
one is a baby boundary and the other is a mother boundary.
The boundaries with small sizes ($\sim {\cal O}(1)$) are called a baby 
one (see the caption of Fig.\ref{fig:3D_Boundary}).
The baby boundaries are originated from the small fluctuations of the 
3D Euclidean spaces.
We thus think that these surfaces are non-universal objects, which 
has been established in 2D QG.
We concentrate on the mother boundary surfaces in this article.
%

\begin{figure}
\centerline{\psfig{file=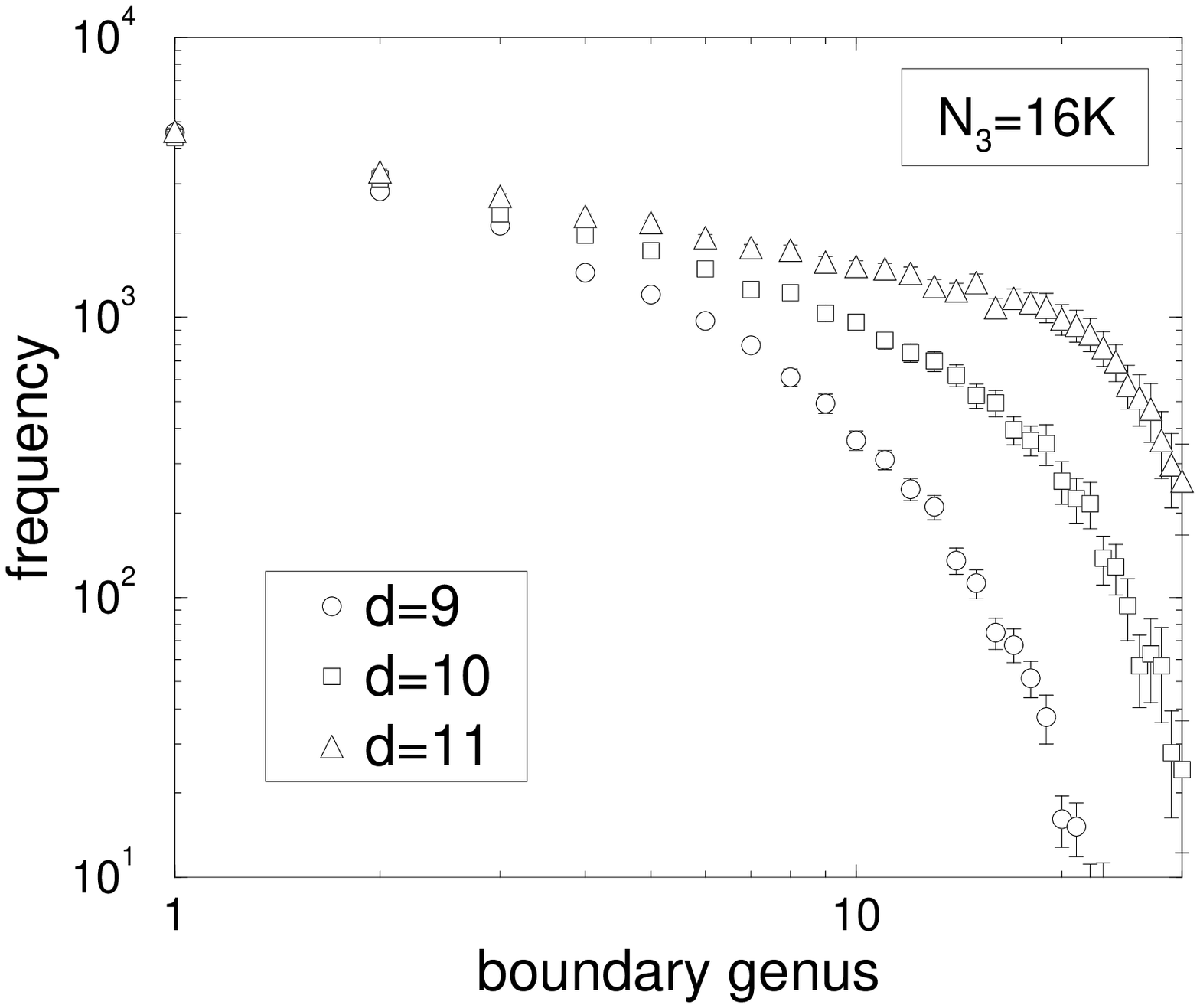,height=6cm,width=6cm}} 
\vspace{-1.0cm}
\caption
{
Genus distributions of the mother boundary near to the critical point with 
log-log scales.
$N_{3}=16K$ ($\kappa_{0}^{c}=4.090$ and $\kappa_{3}=2.200$). 
$d=9,10$ and $11$.
}
\label{fig:BG_Critical_V16K}
\end{figure}

\begin{figure}
\centerline{\psfig{file=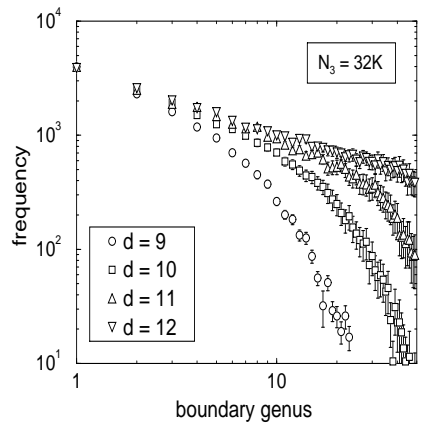,height=6cm,width=6cm}} 
\vspace{-1.0cm}
\caption
{
Genus distributions of the mother boundary near to the critical point with 
log-log scales.
$N_{3}=32K$ ($\kappa_{0}^{c}=4.195$ and $\kappa_{3}=2.222$). 
$d=9,10,11$ and $12$.
}
\label{fig:BG_Critical_V32K}
\end{figure}
\subsection{Genus distributions near to the critical point}
The genus of the boundary surfaces is a dimensionless quantity.
Therefore, it is naively expected that the genus distributions show the 
scaling property; in fact, we have obtained evidence of the scaling relation, 
$P(g) \sim g^{-\alpha}$, where $g$ is a genus.
We can estimate $\alpha \approx 0.5$ from Fig.\ref{fig:BG_Critical_V32K}.
The scaling property of the genus distributions becomes more clear 
the bigger the size of the boundary becomes 
(see Figs.\ref{fig:BG_Critical_V32K} and \ref{fig:BG_Critical_V16K}).
Thus, this scaling property will remain after the thermal limit, $N_{3} 
\to \infty$.
If we go forward into the weak-coupling phase ($\kappa_{0}^{c} < \kappa_{0}$), 
this scaling relation disappears.
However, in the strong-coupling phase we have another type of distribution 
of the genus.

\subsection{Genus distributions in the strong-coupling phase}
\begin{figure}
\centerline{\psfig{file=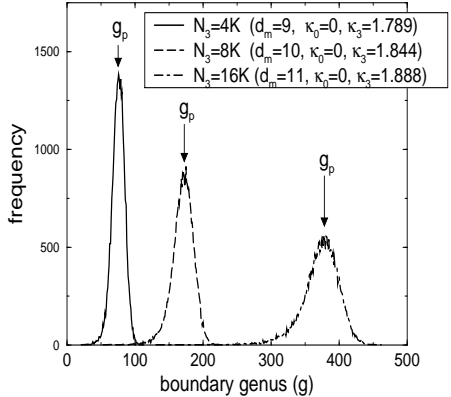,height=6cm,width=6cm}} 
\vspace{-1.0cm}
\caption
{
Genus distributions of the mother boundary surface in the strong-coupling 
limit ($\kappa_{0}=0.0$).
}
\label{fig:BG_Strong}
\end{figure}
We also measure the genus distributions of the mother boundary in 
the strong-coupling limit (i.e., $\kappa_{0}=0$) with volumes $N_{3}=4K,8K$ 
and $16K$ (see Fig.\ref{fig:BG_Strong}).
In this measurement we choose a maximum-distance ($d_{m}$) at which the 
peak value ($g_{p}$) of each distribution becomes maximum, and obtain 
the values $d_{m}=9,10$ and $11$ for $N_{3}=4K,8K$ and $16K$, respectively.
The peak value ($g_{p}$) of each distribution becomes larger the bigger 
the size of boundary surface becomes.
It is well known that the manifold starts to crumple in this phase, and that 
the fractal dimension tends to diverge.
The $g_{p}$ will diverge when $N_{3}$ goes to infinity, which shows that 
the manifold starts to crumple boundlessly in the strong-coupling limit.

\section{Summary and Discussion}
We are concerned with the genus of the mother surfaces appearing on 
boundaries (i.e., sections) of 3D DT mfd with $S^{3}$ topology 
near to the critical point and in the strong-coupling phase.
We find a power-like scaling behavior of genus distributions of the mother 
boundary surfaces near to the critical point: $P(g) \sim g^{-\alpha}$, where 
$\alpha$ is about $0.5$ in our simulation sizes.
Furthermore, in the strong-coupling limit $g_{p}$ seems to diverge when 
$N_{3}$ goes to infinity.

There are some applications of the boundary analysis: (i)the quantum 
nucleation of the universe or topology changing of the boundary 
surfaces in $(2+1)$-dimensional quantum gravity and (ii)the formation of 
nontrivial structures, such as links or knots.

\end{document}